\documentclass[a4paper,12pt]{article}
 \UseRawInputEncoding
\usepackage[left=2.5cm,bottom=3cm,right=2.5cm,top=3cm]{geometry} 
 \usepackage{mathtools}  
\usepackage{makecell}   %
\usepackage{bigints}
\usepackage{overpic,youngtab}
\usepackage{bbm}
\usepackage[left=2.5cm,bottom=3cm,right=2.5cm,top=3cm]{geometry} 

\usepackage{overpic,youngtab}
\usepackage[latin,english]{babel}
\usepackage{amsmath}
\usepackage{amssymb}
\usepackage{epstopdf}
\usepackage{graphics,psfrag,rotating}
\usepackage{mathabx}
\usepackage{graphicx}
\usepackage{dcolumn}
\usepackage{float}
\usepackage{pdflscape}
\usepackage{array}
\usepackage{booktabs}
\usepackage{amscd} 
\usepackage{mathtools}
\usepackage{fancybox}
\usepackage{tensor}
\usepackage{fix-cm}
\usepackage[colorlinks=true,citecolor=blue,,linktocpage=true,linkcolor=blue,urlcolor=black]{hyperref}
\usepackage{tikz} 
\usepackage{cleveref}
\usetikzlibrary{matrix} 
\usetikzlibrary{positioning} 
\usetikzlibrary{arrows}
\usepackage{titlesec}
\usepackage{abstract}
\usepackage{cancel}
\usepackage{pdflscape}

\def\be{\begin{equation}}
\def\ee{\end{equation}}
\def\bea{\begin{eqnarray}}
\def\eea{\end{eqnarray}}
\def\pd{\partial}
\def\a{\alpha}
\def\b{\beta}
\def\g{\gamma}

\def\d{\delta}
\def\m{\mu}
\def\n{\nu}
\def\t{\tau}

\def\l{\lambda}

\def\r{\rho}

\def\bR{\bar{R}}
\def\bD{\bar{D}}
\def\bp{\bar{\phi}}
\def\bn{\bar{\nabla}}
\def\bR{\bar{R}}

\def\s{\sigma}
\def\e{\epsilon}
\def\bi{\begin{itemize}}
	\def\ei{\end{itemize}}
\def\bg{\bar{g}}

\def\bph{\bar{\phi}}

\usepackage[latin,english]{babel}
\usepackage{amsmath}
\usepackage{amssymb}
\usepackage{epstopdf}
\usepackage{graphics,psfrag,rotating}
\usepackage{mathabx}
\usepackage{graphicx}
\usepackage{dcolumn}
\usepackage{float}
\usepackage{pdflscape}
\usepackage{array}
\usepackage{booktabs}
\usepackage{amscd} 
\usepackage{mathtools}
\usepackage{fancybox}
\usepackage{fix-cm}
\usepackage[colorlinks=true,citecolor=blue,,linktocpage=true,linkcolor=blue,urlcolor=black]{hyperref}
\usepackage{tikz} 
\usetikzlibrary{matrix} 
\usetikzlibrary{positioning} 
\usetikzlibrary{arrows}
\usepackage{titlesec}
\usepackage{abstract}
\newcolumntype{M}[1]{>{\centering\arraybackslash}m{#1}}
\newcolumntype{N}{@{}m{0pt}@{}}
\newcommand{\rvline}{\hspace*{-\arraycolsep}\vline\hspace*{-\arraycolsep}}

\def\be{\begin{equation}}
\def\ee{\end{equation}}
\def\bea{\begin{eqnarray}}
\def\eea{\end{eqnarray}}
\def\pd{\partial}
\def\a{\alpha}
\def\b{\beta}
\def\g{\gamma}
\def\xp{x^\prime}
\def\d{\delta}
\def\m{\mu}
\def\n{\nu}
\def\t{\tau}

\def\l{\lambda}

\def\r{\rho}

\def\bR{\bar{R}}
\def\bD{\bar{D}}
\def\bp{\bar{\phi}}
\def\bph{\bar{\phi}}

\def\bn{\bar{\nabla}}
\def\bR{\bar{R}}

\def\bD{\overline{D}}
\def\s{\sigma}

\def\e{\epsilon}
\def\bi{\begin{itemize}}
\def\ei{\end{itemize}}
\def\bg{\bar{g}}

\begin{document}

		\vspace*{-1cm}
		\phantom{hep-ph/***} 
		{\flushleft
			{{FTUAM-19-14}}
			\hfill{{ IFT-UAM/CSIC-19-90}}}
		\vskip 1.5cm
		\begin{center}
		{\LARGE\bfseries Quantum gravity in JNW spacetime.}\\[3mm]
			\vskip .3cm
		
		\end{center}
		\vskip 0.5  cm
		\begin{center}
			{\large Enrique \'Alvarez and  Jes\'us Anero. }
			\\
			\vskip .7cm
			{
				Departamento de F\'{\i}sica Te\'orica and Instituto de F\'{\i}sica Te\'orica, 
				IFT-UAM/CSIC,\\
				Universidad Aut\'onoma de Madrid, Cantoblanco, 28049, Madrid, Spain\\
				\vskip .1cm

				\vskip .5cm
				\begin{minipage}[l]{.9\textwidth}
				\begin{center} 
					\textit{E-mail:} 
					\tt{enrique.alvarez@uam.es},
					\tt{jesusanero@gmail.com} and
					\tt{eduardo.velasco@uam.es}
				\end{center}
			\end{minipage}
			}
		\end{center}
	\thispagestyle{empty}
	
\begin{abstract}\vspace{-1em}
	\noindent
	In this paper we study the behavior of a scalar field coupled to gravitons on the Janis-Newman-Winicour background, which somewhat interpolates between Minkowski and Schwarzschild spacetimes.
	The most important physical effect we find is that there is a 17-dimensional position-dependent { \em mass matrix} $Y_{AB}(x)$ which happens to be  non-diagonal in the basis in which the kinetic energy term is diagonal. There is a different basis  with a mixing between the scalar field and the graviton trace in which the mass matrix is diagonal, but this basis fails to diagonalize the kinetic energy piece.  This is at variance with what happens in the Standard Model with the quark mixing, and is of course due to the fact that the mass matrix here is position dependent and thus it does not commute with the kinetic energy operator, so that both operators cannot be diagonalized simultaneously.

\end{abstract}

\newpage
\tableofcontents
	\thispagestyle{empty}
\flushbottom

\newpage
\setcounter{page}{1}

\section{Introduction  and conclusions.}

Quantum field theory in a (passive) gravitational background has been extensively studied \cite{Birrell}\cite{Hu}. Most physically relevant results on the interaction amongst quantum fields with gravity, like Hawking radiation or the study of the fields assumed to be responsible for inflation, have been obtained in this framework. Although it is known that quantum fields back react on the metric through their energy-momentum tensor, modifying the initial metric  even at the classical level, there are not many explicit computations.
\par
We would like to make a quantitative comparison between this approach, and another one in which the quantum effects on the gravitational  field are taken into account. It is a fact that we do not know how to quantize gravity; this one of the big problems in theoretical physics (cf. for example \cite{EA}). It is possible however to reach a less ambitious goal.
 Standard techniques \cite{Barvinsky} permit the computation of quantum fluctuations around a given spacetime background  at each order in perturbation theory.  Consider the simplest example, namely, the action for an scalar field minimally coupled to the gravitational field reads 

\bea
S_{\text{\tiny{EH$\phi$}}}&&\equiv \int
d^n x\sqrt{|g|}\left[-\frac{1}{2\kappa^2}R+\frac{1}{2}\nabla_\l\phi\nabla^\l\phi+\frac{1}{2}T_{\m\n} h^{\m\n} +\frac{1}{2}J\phi \right]
\eea
where we have included sources for the quantum fields defined as usual by
\bea
&g_{\m\n}\rightarrow\bg_{\m\n}+\kappa h_{\m\n}\nonumber\\
&\phi\rightarrow\bp+\phi
\eea
and the background fields solve the classical equations of motion (EM)
\be
\left.\left.{\d S\over \d \phi}\right|_{g_{\a\b}=\bg_{\a\b},\phi=\bar{\phi}}={\d S\over \d g_{\a\b}}\right|_{g_{\a\b}=\bg_{\a\b},\phi=\bp}=0
\ee
It is a matter of fact  that given an arbitrary Ricci flat gravitational background, and in the absence of any scalar potential, there is always the trivial scalar classical solution, namely $\overline{\phi}=0$. The corresponding energy-momentum tensor vanishes, so that this particular scalar field does not back react on the gravitational field; that is, it does not change the spacetime metric (or Einstein's equations for that matter).
Conversely, as soon as the classical scalar field is nontrivial, the original Ricci flat spacetime cannot be sourced by it, and the spacetime is modified. This is exactly what happens when a nontrivial scalar field is propagating in (for example) Schwarzschild spacetime.
\par
 Our aim in the present paper  is to compute both gravitational and scalar fluctuations with a non-trivial scalar background in a consistent way  to one loop order.  Borrowing from \cite{Barvinsky} the quadratic action after adding  the gauge fixing piece reads
\bea
&&S_{\text{\tiny{2+gf+J}}}=\frac{1}{2}\,\int\,d^nx\sqrt{|g|}\,\bigg\{\,\psi^A\Delta_{AB}\psi^B+ J^A \psi_A\bigg\},
\eea 
where we have written the quadratic operator corresponding to the generalized field, $\psi^A$, defined as a vector
\be
\psi^A\equiv\begin{pmatrix}
	h^{\m\n}\\\phi
\end{pmatrix},
\ee
and the operator takes the (minimal) form in the generalized harmonic gauge
\be
\Delta_{AB}=-g_{AB}\overline{\Box}+Y_{AB}.
\label{op}
\ee
Sources are included for the graviton as well as for the scalar particle
\be
J^A=\begin{pmatrix}T^{\m\n}\\J\end{pmatrix}
\ee

The important thing is that this operator is non-diagonal; there is a mixing between gravitons and scalar particles induced by the term
\be
\left.{\d^2 L\over \d g_{\a\b}\d \phi}\right|_{g_{\a\b}=\bg_{\a\b},\phi=\bar{\phi}} 
\ee
 The components of $g_{AB}$ and $Y_{AB}$ will be  given explicitly later on. This position is dependent mass matrix $Y_{AB}$ is the root of the mixing scalar/graviton in the states that diagonalize it.  As we have just seen, this is due to the universality of the coupling of gravitation to the matter.
 \par

This leads eventually to a free energy in terms of the sources
\be
W\left[J_A\right]=N-{1\over 8} \int J^A (x) \Delta^{-1}_{AB}(x-y) J^B(y)
\ee
because
\be \psi^A\Delta_{AB}\psi^B+\psi^AJ_A=\Big[\psi_A\Delta^{AC}+\frac{1}{2}J^C\Big]\Delta^{-1}_{CD}\Big[\Delta^{DB}\psi_B+\frac{1}{2}J^D\Big]-\frac{1}{4}J^C\Delta^{-1}_{CD}J^D
\ee

In flat space, the propagator 
\be-\Box_0D_0(x-y)=\d(x-y)\ee reads
\be
D_0(x-y)\equiv -\int \frac{d^4 k}{k^2} e^{i k (x-y)}=\frac{1}{(4\pi)^2}\frac{1}{ \left|x-y\right|^2}
\ee
In general the background spacetime, $\bg_{\m\n}$ will not be flat, and we need to know its propagator
\be
-\overline{\Box}  \,\bD(x-y)=\d(x-y)
\ee
Then we can solve the full equation
\be
\left[-\overline{\Box} \d^A_B +Y^A_B(x)\right]D^B_C(x,y)=\d^A_C\d(x-y)
\ee
The full propagator will in general depend on the points $x$ and $y$ separately, and not only on the difference, $x-y$.
We shall define
\be
D^{AB}(x,y)\equiv g^{AC}(x)D_C\,^B(x,y)=D^A\,_C(x,y)g^{CB}(y)
\ee
The solution to the above satisfies the self-consistent integral equation
\be
D^A\,_B(x,y)=\d^A_B\bD(x-y) -\int dz \overline{D}(x-z) Y^A\,_D(z) D^D\,_B(z,y)
\ee
just because
\bea
&&\left(-\tensor{\delta}{^A_B} \Box_x + \tensor{Y}{^A_B}(x) \right)\tensor{D}{^B_C}(x-y)=\left(-\tensor{\delta}{^A_B} \Box_x +\tensor{Y}{^A_B} \right)\Big[\tensor{\delta}{^B_C}\bar{D}(x-y)-\nonumber\\
&&-\int_{}^{} dz  \bar{D}(x-z)\tensor{Y}{^B_D}(z) \tensor{D}{^D_C}(z-y)\Big]=\tensor{\delta}{^A_C}  \delta(x-y)+\nonumber\\
&&+\int_{}^{} dz\Box_x(\bar{D}(x-z))\tensor{Y}{^A_D}(z) \tensor{D}{^D_C}(z-y)+ \tensor{Y}{^A_B}(x) \tensor{D}{^B_C}(x-y)=\tensor{\delta}{^A_C}  \delta(x-y)\nonumber\\
\eea

The former  equation can easily be solved in perturbation theory in the {\em mass} $Y_{AB}$ (what mathematicians call the Neumann series) once we know the fundamental solution of the wave equation in the background spacetime \cite{Friedlander}. To be specific
 \be
D^A\,_B(x,y)=\d^A_B\bD(x-y)-\int dz \overline{D}(x-z) Y^A\,_B(z) \overline{D}(z-y)+\ldots
\ee
\par
 In order to that the first thing we need is a classical background solution of the coupled Einstein-scalar EM. This is not a trivial thing if we postulate a nontrivial scalar background $\overline{\phi}\neq 0$. We emphasize that  the aim of this paper is to contrast such a self-consistent computation with the usual {\em quantum theory in curved spacetime} approach. 
 \par

From the physical point of view, the most important effect is that the 17-dimensional position-dependent { \em mass matrix} $Y_{AB}(x)$ is non-diagonal in the basis in which the kinetic energy term is diagonal. There is a different basis  with a mixing between the scalar field and the graviton trace in which the mass matrix is diagonal, but this basis fails to diagonalize the kinetic energy piece.  This is at variance with what happens in the Standard Model with the quark mixing, and is of course due to the fact that the mass matrix here is position dependent and thus it does not commute with the kinetic energy operator, so that both operators cannot be diagonalized simultaneously.

  Let us discuss in detail what we have  done in the next few paragraphs.

\section{Details on the background expansion}

In order to  be specific,  consider the Einstein-Hilbert action with the metric coupled to a massless scalar field $\phi$ given by
\bea
S_{\text{\tiny{EH$\phi$}}}&&\equiv -\frac{1}{2\kappa^2}\int
d^n x\sqrt{|g|}R+\frac{1}{2}\int
d^n x\sqrt{|g|}\nabla_\l\phi\nabla^\l\phi
\label{ACT} 
\eea

In order to study the effects of quantum gravity for a scalar field, we expand the action to second order, by considering the quantum perturbations, 
\begin{align}
	\phi&\rightarrow \overline{\phi}+  \phi,\nonumber\\
	\tensor{g}{_\mu_\nu} &\rightarrow \tensor{\overline{g}}{_\mu_\nu}+ \kappa \tensor{h}{_\mu_\nu}. 
\end{align}
In all this it is important to bear in mind the dimensionality of the fields, 
\begin{equation}
	\left[\phi\right] =\left[\tensor{h}{_\mu_\nu} \right] =\left[\kappa^{-1}\right] =1.
\end{equation}
the action up to quadratic order in the perturbations reads
\bea 
\label{BF}S_{\text{\tiny{2}}}&&=\int
d^n x~\sqrt{|\bg|}~\left\{h^{\a\b} M_{\a\b\m\n} h^{\m\n}+h^{\a\b}E_{\a\b}\phi+\phi F \phi\right\}.
\eea
the explicit expression for these operators is then
\bea\label{K}
M_{\a\b\m\n}&&= \frac{1}{16}\left(2\bg_{\a\b}\bg_{\m\n}-\bg_{\b\n}\bg_{\a\m}-\bg_{\b\m}\bg_{\a\n}\right)\bar{\Box}-\frac{1}{8}\left(\bg_{\a\b}\bn_\m\bn_\n+\bg_{\m\n}\bn_\a\bn_\b\right)+\nonumber\\
&&+\frac{1}{16}\left(\bg_{\b\n}\bn_\a\bn_\m+\bg_{\b\m}\bn_\a\bn_\n+\bg_{\a\n}\bn_\b\bn_\m+\bg_{\a\m}\bn_\b\bn_\n\right)-\nonumber\\
&&-\frac{1}{16}\left(\bg_{\a\b}\bg_{\m\n}-\bg_{\b\n}\bg_{\a\m}-\bg_{\b\m}\bg_{\a\n}\right)\bR+\frac{1}{8}\left(\bg_{\a\b}\bR_{\m\n}+\bg_{\m\n}\bR_{\a\b}\right)-\nonumber\\
&&-\frac{1}{16}\left(\bg_{\a\m}\bR_{\b\n}+\bg_{\a\n}\bR_{\b\m}+\bg_{\b\m}\bR_{\a\n}+\bg_{\b\n}\bR_{\a\m}\right)-\frac{1}{8}\left(\bR_{\m\a\n\b}+\bR_{\n\a\m\b}\right)\nonumber\\
&&-\frac{\kappa^2}{16}\left(\bg_{\a\m}\bg_{\b\n}+\bg_{\a\n}\bg_{\b\m}-\bg_{\a\b}\bg_{\m\n}\right)\left(\bg^{\r\s}\bn_\r\bph\bn_\s\bph\right)
+\nonumber\\
&&-\frac{\kappa^2}{8}\left(\bg_{\a\b}\bn_\m\bph\bn_\n\bph+\bg_{\m\n}\bn_\a\bph\bn_\b\bph-\bg_{\a\m}\bn_\b\bph\bn_\n\bph-\bg_{\a\n}\bn_\b\bph\bn_\m\bph-\right.\nonumber\\
&&\left.-\bg_{\b\m}\bn_\a\bph\bn_\n\bph-\bg_{\b\n}\bn_\a\bph\bn_\m\bph\right)\nonumber\\
E_{\a\b}&&=\kappa\Big[\frac{1}{2}\bg_{\a\b}\bg^{\r\s}\bn_\r\bph\bn_\s-\frac{1}{2}\bn_\a\bph\bn_\b-\frac{1}{2}\bn_\b\bph\bn_\a\Big]\nonumber\\
F&&=-\frac{1}{2}\bar{\Box}
\eea

Nonetheless, the quadratic action that we have left is still invariant under the quantum gauge symmetry corresponding to diffeomorphism invariance given by the transformations
\bea
\d h_{\m\n}&&=\bn_{\m}\xi_{\n}+\bn_{\n}\xi_{\m}+\mathcal{L}_\xi h_{\m\n}\nonumber\\
\d \phi &&= \xi^\mu \bn_\mu \phi 
\eea
Let us  consequently define the de Donder or harmonic gauge fixing 
\be
S_{\text{\tiny{gf}}}=\frac{1}{4}\,\,\int\,d^nx\,\,\sqrt{\bg}\,\bg_{\m\n}\chi^\m\chi^\n,
\ee
where 
\be
\chi_\n=\bn^\m h_{\m\n}-\frac{1}{2}\bn_\n h-2\kappa\phi\bn_\n\bp.
\ee	
in detail
\bea\label{GF}
S_{\text{\tiny{gf}}}&&=\frac{1}{\xi}\int\,d^nx\,\,\sqrt{\bg}\Bigg\{h^{\a\b} \Big[-\frac{1}{16}\bg_{\a\b}\bg_{\m\n}\bar{\Box}+\frac{1}{8}\left(\bg_{\a\b}\bn_\m\bn_\n+\bg_{\m\n}\bn_\a\bn_\b\right)-\nonumber\\
&&-\frac{1}{16}\left(\bg_{\b\n}\bn_\a\bn_\m+\bg_{\b\m}\bn_\a\bn_\n+\bg_{\a\n}\bn_\b\bn_\m+\bg_{\a\m}\bn_\b\bn_\n\right)\Big] h^{\m\n}+\nonumber\\
&&+\kappa h^{\a\b}\Big[-\frac{1}{2}\bg_{\a\b}\bg^{\r\s}\bn_\r\bph\bn_\s+\frac{1}{2}\bn_\a\bph\bn_\b+\frac{1}{2}\bn_\b\bph\bn_\a+\frac{1}{2}\bn_\a\bn_\b\bph+\frac{1}{2}\bn_\b\bn_\a\bph-\frac{1}{2}\bg_{\a\b}\bar{\Box}\bph\Big]\phi+\nonumber\\
&&+\kappa^2\phi  \Big[\bn_\l\bph\bn^\l\bph\Big]\phi\Bigg\}
\eea
It is then clear that, by setting $\xi=1$ in \eqref{GF} the quadratic operator is a minimal one, so that adding up all the contributions, the quadratic action reads
\bea
S_{\text{\tiny{2+gf}}}=\frac{1}{2}\,\int\,d^nx\,\sqrt{\bg}\,\psi^A\Delta_{AB}\psi^B,
\eea 
where we have written the quadratic operator corresponding to the generalized field, $\psi^A$, defined as the vector
\be
\psi^A\equiv\begin{pmatrix}
	h^{\m\n}\\\phi
\end{pmatrix},
\ee
and the operator takes the form
\be
\Delta_{AB}=-g_{AB}\bar{\Box}+Y_{AB}.
\label{op}
\ee
Let us specify its  different pieces. The term multiplying the box operator acts as a sort of internal metric $g_{AB}$ and reads
\be
g_{AB}=\begin{pmatrix}
	C_{\a\b\m\n}&0\\0&1 \end{pmatrix},
\ee
with 
\be C_{\m\n\r\s}=\frac{1}{8}\left(\bg_{\m\r}\bg_{\n\s}+\bg_{\m\s}\bg_{\n\r}-\bg_{\m\n}\bg_{\r\s}\right)\ee 
defining
\be C^{\m\n\r\s}=2\left(\bg^{\m\r}\bg^{\n\s}+\bg^{\m\s}\bg^{\n\r}-\frac{2}{n-2}\bg^{\m\n}\bg^{\r\s}\right)\ee
then, it is plain that
\be
g^{AB}=\begin{pmatrix}
	C^{\a\b\m\n}&0\\0&1 \end{pmatrix},
\ee
implies
\be g_{AB}g^{BC}=\d_A^C\ee
and finally the components of $Y_{AB}$
\bea
&&Y^{hh}_{AB}=\frac{1}{8}\left(\bg_{\b\n}\bg_{\a\m}+\bg_{\b\m}\bg_{\a\n}-\bg_{\a\b}\bg_{\m\n}\right)\bR+\frac{1}{4}\left(\bg_{\a\b}\bR_{\m\n}+\bg_{\m\n}\bR_{\a\b}\right)-\nonumber\\
&&-\frac{1}{8}\left(\bg_{\a\m}\bR_{\b\n}+\bg_{\a\n}\bR_{\b\m}+\bg_{\b\m}\bR_{\a\n}+\bg_{\b\n}\bR_{\a\m}\right)-\frac{1}{4}\left(\bR_{\m\a\n\b}+\bR_{\n\a\m\b}\right)-\nonumber\\
&&-\frac{\kappa^2}{8}(\bg_{\m\r}\bg_{\n\s}+\bg_{\m\s}\bg_{\n\r}-\bg_{\m\n}\bg_{\r\s})\bg^{\a\b}\bn_\a\bph\bn_\b\bph+\nonumber\\
&&+\frac{\kappa^2}{4}\left(\bg_{\m\a}\bn_\n\bph\bn_\b\bph+\bg_{\m\b}\bn_\n\bph\bn_\a\bph+\bg_{\n\a}\bn_\m\bph\bn_\b\bph+\bg_{\n\b}\bn_\m\bph\bn_\a\bph-\bg_{\m\n}\bn_\a\bph\bn_\b\bph-\bg_{\a\b}\bn_\m\bph\bn_\n\bph\right)\nonumber\\
&&Y^{h\phi}_{AB}=Y^{\phi h}_{AB}=\kappa\Big[
\frac{1}{2}\left(\bn_\a\bn_\b\bph+\bn_\b\bn_\a\bph\right)-\frac{1}{2}\bg_{\a\b}\bar{\Box}\bph\Big]\nonumber\\
&&Y^{\phi\phi}_{AB}=2\kappa^2\bg^{\r\s}\bn_\r\bph\bn_\s\bph\eea
simbolically, the operation, $\psi^A \tensor{Y}{_A^B} \psi_B= \psi_A \tensor{Y}{^A^B} \psi_B$, can be represented by, 
\begin{align}
	&\begin{array}{cc}
		(\tensor{h}{_\alpha_\beta} & \phi )
	\end{array}		
	\begin{pmatrix}
		\begin{matrix}
		\tensor{Y}{^\alpha^\beta^\gamma^\delta_{hh}} 
		\end{matrix}
		& \rvline & \tensor{Y}{^\gamma^\delta_{\phi h }}  \\
	  \hline
	  \tensor{Y}{^\alpha^\beta_{ h \phi}} & \rvline &
		\begin{matrix}
		Y_{\phi \phi}
		\end{matrix}
	  \end{pmatrix}
	  \left(\begin{array}{c}
		\tensor{h}{_\gamma_\delta} \\
		\phi
	\end{array}\right)=\nonumber\\
	&=\tensor{h}{_\alpha_\beta}\tensor{Y}{^\alpha^\beta^\gamma^\delta_{hh}}\tensor{h}{_\gamma_\delta}+ \tensor{h}{_\alpha_\beta} \tensor{Y}{^\alpha^\beta_{ h \phi}} \phi +\phi \tensor{Y}{^\gamma^\delta_{\phi h }} \tensor{h}{_\gamma_\delta} + \phi Y_{\phi \phi} \phi \label{207} 
\end{align}		
By inspection of \eqref{207} it is clear that upon a field redefinition of the scalar field, 
\begin{equation}
	\varphi=\phi +  \frac{\tensor{Y}{^\alpha^\beta_{ h \phi}}}{ Y_{\phi \phi} } \tensor{h}{_\alpha_\beta}\label{varfi} 
\end{equation}
we can write \cref{207} as, 
	\begin{align}
		&\begin{array}{cc}
			(\tensor{h}{_\alpha_\beta} & \varphi )
		\end{array}		
		\begin{pmatrix}
			\begin{matrix}
			\tensor{\hat{Y}}{^\alpha^\beta^\gamma^\delta_{hh}} 
			\end{matrix}
			& \rvline & 0 \\
		  \hline
		 0 & \rvline &
			\begin{matrix}
			\hat{Y}_{\varphi \varphi}
			\end{matrix}
		  \end{pmatrix}
		  \left(\begin{array}{c}
			\tensor{h}{_\gamma_\delta} \\
			\varphi
		\end{array}\right)=\nonumber\\
		&=\tensor{h}{_\alpha_\beta}\tensor{\hat{Y}}{^\alpha^\beta^\gamma^\delta_{hh}}\tensor{h}{_\gamma_\delta}+\varphi \hat{Y}_{\varphi \varphi} \varphi 
	\end{align}	
	where the elements of the block diagonal matrix are related to the original ones by, 
	\begin{align}
		\tensor{\hat{Y}}{^\alpha^\beta^\gamma^\delta_{hh}}&=\tensor{Y}{^\alpha^\beta^\gamma^\delta_{hh}}-\frac{\tensor{Y}{^\alpha^\beta_{ h \phi}}\tensor{Y}{^\gamma^\delta_{ h \phi}}}{Y_{\phi \phi} } \nonumber\\
		\hat{Y}_{\varphi \varphi}&=Y_{\phi \phi}
	\end{align}
	Note that this kind of block diagonalization is highly non-unique as any matrix that only mixes the gravitational or the \textit{scalar-dressed} sector gives a matrix similar to the one above. 
\par 
In conclusion, we have found the basis of operators that diagonalize the position dependent mass matrix; this is just $h_{\m\n},\varphi$, that is, there is some mixing between the scalar field and the  graviton field. As we already observed in the Introduction, the fact that the mass matrix does not commute with the kinetic energy operator (precisely because it is position dependent; as a matter of fact it is not really a mass matrix except in a loose sense) prevents us from diagonalizing them simultaneously.
\section{The classical background:  JNW spacetime.}
There is a very interesting static and spherically symmetric exact solution of the  massless scalar field coupled to gravity discovered by Janis, Newman and Winicour \cite{Janis}\cite{Virbhadra}\cite{Wyman}. This is the simplest consistent background solution that we have found, and it is very important because it somewhat interpolates between Minkowski  and Schwarzschild spacetimes .
\par

The JNW spacetime is a particular case of the Plebanski-Demianski \cite{Plebanski} spacetime, which is the most general type D Ricci flat spacetime. As such, they are related to some self-dual two form by the mechanism of {\em Weyl doubling} \cite{Alawadhi}. This is reviewed in the appendix.
\par

\bea ds^2&&=\left(1-\frac{b}{\r}\right)^\g\text{dt}^2-\left(1-\frac{b}{\r}\right)^{-\g}\text{d$\r$}^2-\left(1-\frac{b}{\r}\right)^{1-\g}\r^2\text{d$\Omega$}^2\nonumber\\
&&\phi={\sqrt{1-\g^2\over 2\kappa^2}}\log\left(1-{b\over \r}\right)
\eea
where the variable $\r\in (b,+\infty)$ and $b$ has the relationship with the Schwarzschild radius $b=\frac{r_s}{\g}$

The components of $g_{AB}$ are
\bea
&&C^{hh}_{tttt}=\frac{1}{8}\left(1-\frac{b}{\r}\right)^{2\g}\nonumber\\
&&C^{hh}_{t\r t\r}=C^{hh}_{t\r\r t}=-\frac{1}{8}\nonumber\\
&&C^{hh}_{t\theta t\theta}=C^{hh}_{t\theta\theta t}=\frac{1}{8}\r(b-\r)\nonumber\\
&&C^{hh}_{\r\r\r\r}=\frac{1}{8}\left(1-\frac{b}{\r}\right)^{-2\g}\nonumber\\
&&C^{hh}_{\r\theta \r\theta}=C^{hh}_{\r\theta \theta \r}=\frac{1}{8}\r^2\left(1-\frac{b}{\r}\right)^{1-2\g}\nonumber\\
&&C^{hh}_{\theta\theta\theta\theta}=\frac{1}{8}\r^4\left(1-\frac{b}{\r}\right)^{2-2\g}\nonumber\\
&&C^{hh}_{\theta\varphi\theta\varphi}=C^{hh}_{\varphi\theta\varphi\theta}=C^{hh}_{\theta\varphi\varphi\theta}=\frac{1}{8}\r^4\left(1-\frac{b}{\r}\right)^{2-2\g}\sin^2\theta\nonumber\\
&&C^{hh}_{\varphi\varphi\varphi\varphi}=C^{hh}_{\theta\theta\theta\theta}\sin^4\theta\nonumber\\
&&C^{hh}_{tt\r\r}=\frac{1}{8}\nonumber\\
&&C^{hh}_{tt\theta\theta}=-\frac{1}{8}\r(b-\r)\nonumber\\
&&C^{hh}_{\r\r\theta\theta}=\frac{1}{8}\r(b-\r)\left(1-\frac{b}{\r}\right)^{-2\g}\nonumber\\
&&C^{hh}_{\theta\theta\varphi\varphi}=-\frac{1}{8}\r^4\left(1-\frac{b}{\r}\right)^{2-2\g}\sin^2\theta
\eea
Obviously $C^{\phi h}=0$ and $C^{\phi\phi}=1$ and the components of $Y_{AB}$ yields
\begin{align}
	Y^{\phi \phi}&=\frac{(\g^2-1)}{\r^2}\left(1-\frac{b}{\r}\right)^{\g-2}\frac{b^2}{\r^2}
\end{align}
\begin{align}
	Y^{h \phi}_{tt}&=-\frac{ \g^2b^2 }{2 \r^4}\left(1-\frac{b}{\r}\right)^{2\g-2}\sqrt{\frac{1}{2}\left(\frac{1}{\g^2}-1\right)}\nonumber\\
	Y^{h \phi}_{\r\r}&=\frac{  b\g \left[-4 \r+b(\g+2)\right]}{2 \r^4}\left(1-\frac{b}{\r}\right)^{-2}\sqrt{\frac{1}{2}\left(\frac{1}{\g^2}-1\right)}\nonumber\\
	Y^{h \phi}_{\theta  \theta}&=\frac{  b\g \left[-2 \r+b(\g+1)\right]}{2 \r^2}\left(1-\frac{b}{\r}\right)^{-1}\sqrt{\frac{1}{2}\left(\frac{1}{\g^2}-1\right)}
\end{align}
\begin{align}
	Y^{h h}_{t \r t \r}&=Y^{h h}_{t \r \r t }=-\frac{  b \left[-4\g \r+b(3\g^2+2\g-1)\right]}{16 \r^4}\left(1-\frac{b}{\r}\right)^{\g-2}\nonumber\\
	Y^{h h}_{t \theta t \theta}&=Y^{h h}_{t \theta  \theta t }=-\frac{  b\g \left[-2 \r+b(\g+1)\right]}{16 \r^2}\left(1-\frac{b}{\r}\right)^{\g-1}\nonumber\\
	Y^{h h}_{\r \r \r \r}&=-\frac{\left(1-\g ^2\right) b^2 }{4 \r^4 }\left(1-\frac{b}{\r}\right)^{-\g -2}\nonumber\\
	Y^{h h}_{\r \theta \r \theta}&=Y^{h h}_{\r \theta  \theta \r}=-\frac{  b \left[2\g \r +b(\g^2-\g-2)\right]}{16 \r^2}\left(1-\frac{b}{\r}\right)^{-\g-1}\nonumber\\
	Y^{h h}_{\theta \varphi\theta \varphi}&=Y^{h h}_{\varphi\theta \varphi \theta}=Y^{h h}_{\theta\varphi \varphi \theta}=\frac{  b \left[-4\g \r +b(\g+1)^2\right]}{16}\left(1-\frac{b}{\r}\right)^{-\g} \sin ^2(\theta ) \nonumber\\
	Y^{h h}_{t t \r \r}&=\frac{  b\g \left[-2 \r +b(\g+1)\right]}{4\r^4}\left(1-\frac{b}{\r}\right)^{\g-2}\nonumber\\
	Y^{h h}_{t t \theta \theta}&=\frac{  b\g \left[-2 \r +b(\g+1)\right]}{8\r^2}\left(1-\frac{b}{\r}\right)^{\g-1}\nonumber\\
	Y^{h h}_{r r \theta \theta}&=-\frac{  b \left[-2\g \r +b(\g+1)\right]}{8\r^2}\left(1-\frac{b}{\r}\right)^{-\g-1}\nonumber \\
	Y^{h h}_{\theta \theta \phi \phi}&=-\frac{  b \left[-4\g \r +b(\g+1)^2\right]}{8}\left(1-\frac{b}{\r}\right)^{-\g} \sin ^2(\theta ) \nonumber\\
\end{align} 
The rest of the components are either vanishing or related by symmetries, recall the matriz has the symmetries $(\alpha\leftrightarrow \beta),(\gamma\leftrightarrow \delta),(\alpha\,\beta)\leftrightarrow(\gamma\, \delta )$. Along with the spherical symmetry relating $\theta $ and  $ \phi$ components.

\section{One-loop divergences}
Let us first clarify first an important issue on divergences in the partition function.
We have seen that up to quadratic terms in the quantum fluctuations,
\bea
S_{\text{\tiny{2+gf}}}=\frac{1}{2}\,\int\,d^nx\,\sqrt{\bg}\,\psi^A\Delta_{AB}\psi^B,
\eea 
where 
\be
\psi^A\equiv\begin{pmatrix}
	h^{\m\n}\\\phi
\end{pmatrix},
\ee
and
\be
\Delta_{AB}=-g_{AB}\bar{\Box}+Y_{AB}.
\label{op}
\ee
The complete path integral
\be
Z[J]\equiv e^{-W[J]}\equiv  \int {\cal D} \psi^A\, e^{-\frac{1}{2}\,\int d(vol)\left\{\,\psi^A\Delta_{AB}\psi^B+ J^A \psi_A\right\}}=e^{-{1\over 2}\text{tr}\log\Delta_{AB}-{1\over 4 }\int d(vol) J^A \Delta^{-1}_{AB} J^B}
\ee
All one-loop divergences are contained in the first term of the exponent, which is independent of the sources. In the absence of scalar self interaction, which is what we are assuming, Green functions,  defined as
\be
{\d^k\left(W[J]-W[0]\right)\over \d J_{A_1}(x_1)\ldots \d J_{A_k}(x_k)}\equiv \langle \text{out}\left|\psi_{A_1}(x_1)\ldots\psi_{A_n}(x_n)\right| \text{in} \rangle
\ee
are independent of the renormalization of the determinant of $\Delta_{AB}$, which at any rate can be easily performed using standard techniques \cite{Barvinsky}\cite{Birrell}.
\par
\section{The massless scalar propagator far from from the singularity}
Let us then proceed by expanding the metric in powers of $\frac{b}{\r}$. Large distances here will then mean $\e_1\ll1$. In this way we avoid any concerns on the naked singularity at $ \r=b$.

The JNW metric up to second order in $\frac{b}{\r}$
\bea &&ds^2=\left(1-\frac{b\g}{\r}-\frac{\g(1-\g) b^2}{2\r^2}\right)\text{dt}^2-\left(1+\frac{b\g}{\r}+\frac{\g(1+\g)b^2}{2\r^2}\right)\text{d$\r$}^2-\nonumber\\
&&-\r^2\left(1-\frac{(1-\g)b}{\r}-\frac{\g(1-\g)b^2}{2\r^2}\right)\text{d$\Omega$}^2\nonumber\\
\eea
and the scalar field
\bea
\phi[\r]&&=-\sqrt{\frac{1-\g^2}{2\kappa^2}}\left[\frac{ b}{\r }+\frac{ b^2}{2\r^2}\right]
\eea
In this metric the components of $g_{AB}$ at second order in $\frac{b}{\r}$ reads
\bea\label{C}
&&C^{hh}_{tttt}=\frac{1}{8}\Big[1-\frac{2\g b}{\r}+\frac{\g(2\g-1)b^2}{\r^2}\Big]\nonumber\\
&&C^{hh}_{t\r t\r}=C^{hh}_{t\r\r t}=-\frac{1}{8}\nonumber\\
&&C^{hh}_{t\theta t\theta}=C^{hh}_{t\theta\theta t}=-\frac{\r^2}{8}\Big[1-\frac{2b}{\r}\Big]\nonumber\\
&&C^{hh}_{\r\r\r\r}=\frac{1}{8}\Big[1+\frac{2\g b}{\r}+\frac{\g(1+2\g)b^2}{\r^2}\Big]\nonumber\\
&&C^{hh}_{\r\theta \r\theta}=C_{\r\theta \theta \r}=\frac{\r^2}{8}\Big[1+\frac{(2\g-1)b}{\r}+\frac{\g(2\g-1)b^2}{\r^2}\Big]\nonumber\\
&&C^{hh}_{\theta\theta\theta\theta}=\frac{\r^4}{8}\Big[1+\frac{2(\g-1)b}{\r}+\frac{(\g-1)(2\g-1)b^2}{\r^2}\Big]\nonumber\\
&&C^{hh}_{\theta\varphi\theta\varphi}=C^{hh}_{\varphi\theta\varphi\theta}=C^{hh}_{\theta\varphi\varphi\theta}=\frac{\r^4}{8}\Big[1+\frac{2(\g-1)b}{\r}+\frac{(\g-1)(2\g-1)b^2}{\r^2}\Big]\sin^2\theta\nonumber\\
&&C^{hh}_{\varphi\varphi\varphi\varphi}=C^{hh}_{\theta\theta\theta\theta}\sin^4\theta\nonumber\\
&&C^{hh}_{tt\r\r}=\frac{1}{8}\nonumber\\
&&C^{hh}_{tt\theta\theta}=\frac{\r^2}{8}\Big[1-\frac{2b}{\r}\Big]\nonumber\\
&&C^{hh}_{\r\r\theta\theta}=-\frac{\r^2}{8}\Big[1+\frac{(2\g-1)b}{\r}+\frac{\g(2\g-1)b^2}{\r^2}\Big]\nonumber\\
&&C^{hh}_{\theta\theta\varphi\varphi}=-\frac{\r^4}{8}\Big[1+\frac{2(\g-1)b}{\r}+\frac{(\g-1)(2\g-1)b^2}{\r^2}\Big]\sin^2\theta
\eea
obviously $C^{\phi h}=0$ and $C^{\phi\phi}=1$.

The components of the mass matrix $Y_{AB}$ at second order in $\frac{b}{\r}$ yields
\begin{align}\label{Y1}
	Y^{\phi \phi}&=\frac{(\g^2-1) }{\r^2}\frac{b^2}{\r^2}
\end{align}
\begin{align}\label{Y2}
	Y^{h \phi}_{tt}&=-\frac{ \g^2 }{2 \r^2}\sqrt{\frac{1}{2}\left(\frac{1}{\g^2}-1\right)}\frac{b^2}{\r^2}\nonumber\\
	Y^{h \phi}_{\r\r}&=\Big[-\frac{2  \g b}{ \r^3}+\frac{ \g(\g-6)b^2}{2\r^4}\Big]\sqrt{\frac{1}{2}\left(\frac{1}{\g^2}-1\right)}\nonumber\\
	Y^{h \phi}_{\theta  \theta}&=\Big[\frac{ \g b}{ \r }-\frac{\g(\g-1)b^2}{2\r^2}\Big]\sqrt{\frac{1}{2}\left(\frac{1}{\g^2}-1\right)}
\end{align}
\begin{align}\label{Y3}
	Y^{h h}_{t \r t \r}&=Y^{h h}_{t \r \r t }=\frac{\g b}{4 \r^3 }+\frac{\left[1+(6-7\g)\g\right]b^2}{16\r^4}\nonumber\\
	Y^{h h}_{t \theta t \theta}&=Y^{h h}_{t \theta  \theta t }=-\frac{\g b }{8 \r } -\frac{\g(3\g-1)b^2}{16\r^2}\nonumber\\
	Y^{h h}_{\r \r \r \r}&=\frac{\left(\g ^2-1\right) b^2 }{4 \r^4 }\nonumber\\
	Y^{h h}_{\r \theta \r \theta}&=Y^{h h}_{\r \theta  \theta \r}=\frac{\g b}{8 \r }+\frac{(3\g^2+\g-2)b^2}{16\r^2}\nonumber\\
	Y^{h h}_{\theta \phi\theta \phi}&=Y^{h h}_{\phi\theta \phi \theta}=Y^{h h}_{\theta\phi \phi \theta}=\r^2\Big[-\frac{\g b}{4\r}+\frac{\left[1+\g(2-3\g)\right]b^2}{16\r^2} \Big]\sin ^2(\theta )\nonumber\\
	Y^{h h}_{t t \r \r}&=-\frac{\g b }{2 \r^3 }+\frac{3\g(\g-1)b^2}{4\r^4}\nonumber\\
	Y^{h h}_{t t \theta \theta}&=\frac{\g b}{4 \r }+\frac{\g(1-3\g)b^2}{8\r^2}\nonumber\\
	Y^{h h}_{\r \r \theta \theta}&=-\frac{\g b}{4 \r }-\frac{(2\g-1)(\g+1)b^2}{8\r^2}\nonumber \\
	Y^{h h}_{\theta \theta \varphi \varphi}&=\r^2\Big[\frac{\g b}{2}+\frac{\left[(\g-1)(1+3\g)\right]b^2}{8}\Big]\sin ^2(\theta )\nonumber\\
\end{align} 
the rest of the components are either vanishing or related by symmetries, recall the matriz has the symmetries $(\alpha\leftrightarrow \beta),(\gamma\leftrightarrow \delta),(\alpha\,\beta)\leftrightarrow(\gamma\, \delta )$. Along with the spherical symmetry relating $\theta $ and  $ \varphi$ components. 

The propagator obeys the equation
\be
\left(-\overline{\Box}\d^A_C +Y^A_C\right) D^C_B(x-\xp)=\d^A_B \d (x-\xp)
\ee
where $\bar{\Box}$ is the d\' Alembertian corresponding to the full JNW metric. This is a formidable equation, which we are unable to solve  exactly. We shall do it in perturbation theory, in two steps.
We first write a solution as a  Liouville-Newmann series
\begin{equation}
\overline{D}_{CB}(x,x'):= g_{CB}\bar{D}(x,x')-\int_{}^{} d^n z  \bar{D}(x,z)Y_{CB}(z) \bar{D}(z,x')
\end{equation}
where 
\be\label{D}
-\overline{\Box} \overline{D}(x-\xp)=\d(x-\xp)
\ee
The second step is to solve this equation in power series of ${b\over \r}$, taking into account that the first term in the series is the flat space propagator, expressed in polar coordinates $(t,\r,\theta,\phi)$
\bea
&&\overline{\Box}=-\Box_0+{b\over \r}\overline{\Box}_1+\ldots\nonumber\\
&&\overline{D}(x-\xp)=D_0(x-\xp)+{b\over \r} \overline{D}_1(x-\xp)+\ldots
\eea
Here 
\bea
&&\overline{\Box}_{(0)}={\pd^2\over \pd t^2}-{\pd^2\over \pd \r^2}-{2\over \r}{\pd \over \pd \r}-{\cot\,\theta\over \r^2}{\pd \over \pd \theta}-{1\over \r^2}{\pd^2\over \pd \theta^2}-{1\over \r^2 \sin^2 \theta}{\pd^2\over \pd \phi^2}\nonumber\\
&&\overline{\Box}_{(1)}=\g{\pd^2\over \pd t^2}+\g {\pd^2\over \pd \r^2}-\frac{(1-\g)}{\r^2}\Big[\cot\,\theta{\pd \over \pd \theta}+{\pd^2\over \pd \theta^2}+{1\over  \sin^2 \theta}{\pd^2\over \pd \phi^2}\Big]+{2\g-1\over \r}{\pd \over \pd \r}\nonumber\\
\eea
and 
\bea
&&D_0(x_1-x_2)=\frac{1}{(4\pi)^2}{1\over |x-\xp|^2}=\nonumber\\
&&=\frac{1}{(4\pi)^2}{1\over |(t_1-t_2)^2-r_1^2-r_2^2+2r_1r_2\left[\cos\theta_1\cos\theta_2+\sin\theta_1\sin\theta_2\cos(\phi_1-\phi_2)\right]|}
\eea

The equation \eqref{D} leads immediatly to

\be \bar{D}_{(1)}(x_1,x_2)=-\int dz  \overline{D}_{(0)}(x_1-z) \overline{\Box}^{(1)}_z \overline{D}_{(0)}(z-x_2)\ee

where
\bea
&&\overline{\Box}^{(1)}_x \overline{D}_{(0)}(x-x_2)=\frac{-1}{r^2 |x-x_2|^6}\Bigg\{\nonumber\\
&&r (1 - 2 \gamma) \Bigl(2 r - 2 r_2 \bigl(\cos(\theta) \cos(\theta_2) + \cos(\phi -  \phi_2) \sin(\theta) \sin(\theta_2)\bigr)\Bigr)|x-x_2|^2-\nonumber\\&&- 2 r^2 \gamma \Bigl(r^2 + r_2^2 + 3 (t -  t_2)^2 - 2 r r_2 \bigl(\cos(\theta) \cos(\theta_2) + \cos(\phi -  \phi_2) \sin(\theta) \sin(\theta_2)\bigr)\Bigr) -\nonumber\\
&&- 2 r^2 \gamma \Bigl(|x-x_2|^2  + 4 \bigl(- r + r_2 \cos(\theta) \cos(\theta_2) + r_2 \cos(\phi -  \phi_2) \sin(\theta) \sin(\theta_2)\bigr)^2\Bigr) +\nonumber\\
&&+ 2 r_2 (1 -  \gamma) \Biggl(4 r \bigl(\cos(\theta_2) \sin(\theta) -  \cos(\theta) \cos(\phi -  \phi_2) \sin(\theta_2)\bigr) \times\nonumber\\
&&\times\bigl(r -  r_2 \cos(\theta) \cos(\theta_2) -  r_2 \cos(\phi -  \phi_2) \sin(\theta) \sin(\theta_2)\bigr)+ \nonumber \\ 
&& +\bigl(\cos(\theta_2) \sin(\theta) -  \cos(\theta) \cos(\phi -  \phi_2) \sin(\theta_2)\bigr)|x-x_2|^2+ \nonumber \\ 
&& +r \cot(\theta) \bigl(\cos(\theta_2) \sin(\theta) -  \cos(\theta) \cos(\phi -  \phi_2) \sin(\theta_2)\bigr) |x-x_2|^2 +\nonumber\\
&&+  r \csc(\theta) \sin(\theta_2) \biggl(\cos(\phi -  \phi_2) |x-x_2|^2 + 4 r r_2 \sin(\theta) \sin(\theta_2) \bigl(\sin(\phi -  \phi_2)\bigr)^2\biggr)\Biggr)\Bigg\}
\eea 
The final answer for the propagator is then the somewhat convoluted equation, at first order in $\frac{b}{\r}$
\bea
&&D^{(1)}_{AB}(x_1,x_2)=g^{(0)}_{AB}\bar{D}_{(1)}(x_1,x_2)+g^{(1)}_{AB}\bar{D}_0(x_1,x_2)-\int_{}^{} dz  \bar{D}_0(x_1,z)Y^{(1)}_{AB}(z) \bar{D}_0(z,x_2)=\nonumber\\
&&=-g^{(0)}_{AB}\int dz  \overline{D}_{(0)}(x_1-z) \overline{\Box}^{(1)}_z \overline{D}_{(0)}(z-x_2)+g^{(1)}_{AB}\bar{D}_0(x_1,x_2)-\int_{}^{} dz  \bar{D}_0(x_1,z)Y^{(1)}_{AB}(z) \bar{D}_0(z,x_2)\nonumber\\
\eea
(The expansion of the metric $g^{(0)}_{AB}$ and $g^{(1)}_{AB}$ has been obtained in \eqref{C}, and the one of the mass matrix $Y^{(1)}_{AB}$ in \eqref{Y1}, \eqref{Y2} and \eqref{Y3})
\subsection{Singularities of the propagator.}
	The two most important properties of the propagator is the position of the poles, and the corresponding residues. Let us examine how this work in a simplified setting.
	Consider the Klein-Gordon operator ${\cal O}$ given in the physical signature by
	\be
	{\cal O}\equiv -\left( \Box+m^2\right),
	\ee
	with constant $m$. 
	The corresponding propagator $D_m(x-\xp)$ is the massive Klein-Gordon propagator and is well known and its divergences can be computed from the short proper time expansion of the heat kernel in one fell swoop. 
	\par
	But we could also define it as a formal series in  the mass parameter $m^2$ and the massless propagator $D_0(x-\xp)$ using the formal series expansion 
	\bea
&&	D_m(x-\xp)\equiv {1\over \Box+m^2}={1\over \Box\left(1+\Box^{-1} m^2\right)}={1\over \Box}\,\sum_{p=0}^\infty (\Box^{-1} m^2)^p=\nonumber\\
&&=D_0(x-\xp) \sum_{p=0}^\infty \left(m^2 D_0(x-\xp) \right)^p
	\eea
	Where the products of propagators are to be interpreted as convolutions.
	This expansion obeys a self-consistent equation
	\be
	D_m(x-\xp)=D_0(x-\xp)+m^2 \int d^n z\,D_0(x-z)D_m(z-\xp)
	\ee
	Which resembles somewhat our integral equation. The question is whether we  get the same physical answers using both methods. Let us first examine the divergent piece.
	
	The heat kernel coefficients can be found in the literature so that the divergent piece (in $n=4$) of the operator reads
	\begin{eqnarray}
	\frac{1}{2}\log det\Delta&=& \frac{1}{n-4}\frac{1}{(4\pi)^{n/2}}\left[ -\frac{1}{6}\bar{R}m^2+\frac{1}{2} m^4 +\frac{1}{360}\left(5\bar{R}^2-2\bar{R}_{\m\n}^2+2\bar{R}_{\m\n\rho\sigma}^2\right) \right]
	\end{eqnarray}
	Looking up  the literature we check that what he calls $a_2$ has in it some mass terms, namely
	\be
	e_4={1\over 360(4\pi)^{n/2}}\left\{-60 R m^2+180 m^4+\ldots\right\}
	\ee
	There is however another way of computing the same divergent piece of the determinant, namely, integrating the mass independently 
	\be
	K(m)=K(0) e^{-m^2\t}={1\over (4\pi\t)^{n/2}}\,\sum_p e_p(-\Box-m^2) \t^{p\over 2}
	\ee
	with
	\be
	\text{tr}\, K(m=0)={1\over (4\pi\t)^{n/2}}\,\sum_p e_p(-\Box) \t^{p\over 2}
	\ee
	This yields
	\bea
	&&K(m)=K(0) e^{-m^2\t}={1\over (4\pi\t)^{n/2}}\,\sum_p e_p(-\Box) \t^{p\over 2}\sum_q {\left(-m^2 \t\right)^q\over q!}=\nonumber\\
	&&={1\over (4\pi\t)^{n/2}}\,\bigg\{e_0(-\Box)\left(1 -m^2\t+{m^4 \t^2\over 2}+\ldots\right)+e_2 (-\Box)\t\left(1-m^2\t+\ldots\right)+\nonumber\\
	&&+e_4(-\Box)\t^2\left(1+\ldots\right)+\ldots\bigg\}
	\eea
	because
	\bea \frac{1}{2}\log det\Delta&&=-\frac{1}{2}\int\frac{d\t}{\t}\frac{1}{(4\pi\t)^{n/2}}e^{-m^2\t}\sum_{p=0}^{\infty}a_p\t^p=-\frac{1}{(4\pi)^{n/2}}\sum_{p=0}^{\infty}a_p \, m^{n-2p} \, \Gamma\left(p-\frac{n}{2}\right),\nonumber\\\eea
	so that the all the mass dependence is treated exactly. In $n=4-\e$ we have
	\bea \frac{1}{2}\log det\Delta&&=\frac{1}{n-4}\frac{1}{(4\pi)^2}\left[a_2(\bar{-\Box})-m^2a_1(\bar{-\Box})+\frac{1}{2}m^4a_0(\bar{-\Box})\right]\eea
	The difference here is that $a_p(-\Box)$ is independent of $m$ and taking the values of the various heat kernel coefficients from the literature we get
	\bea
		a_0(\bar{-\Box})&&=1\nonumber\\
		a_1(\bar{-\Box})&&=\frac{1}{6}\bR\nonumber\\
		a_2(\bar{-\Box})&&=\frac{1}{360}\left(5\bar{R}^2-2\bar{R}_{\m\n}^2+2\bar{R}_{\m\n\rho\sigma}^2\right). \eea
	We see that we indeed obtain the same result using both methods. Expanding the above
	\bea
	&&e_4(-(\Box+m^2))={1\over (4\pi)^{n/2}}\Big[e_4(-\Box)-e_2(-\Box) m^2+ {m^4\over 2}e_0(-\Box)\Big]=\nonumber\\
	&&{1\over (4\pi)^{n/2}}\bigg\{{1\over 360 }\Big[12 \Box R+ 5 R^2 -2 R_{\m\n}^2+ 5R_{\m\n\r\s}^2\Big]- {m^2\over 6}\,R+{m^4\over 2}\bigg\}
	\eea
	because 
	\bea
	e_0(-\Box)&&=1\nonumber\\
		e_2(-\Box)&&={1\over 6} R\nonumber\\
	e_4(-\Box)&&={1\over 360 }\bigg\{12 \Box R+ 5 R^2 -2 R_{\m\n}^2+2 R_{\m\n\r\s}^2\bigg\}
	\eea

Let us now examine the singularities. The massive propagator is given by
	\bea
	&&\Delta(x-x^\prime)=\int d\t K(x,x^\prime;\t)=\int_0^\infty d\t\,{1\over (4\pi\t)^{n/2}}\,e^{-{(x-x^\prime)^2\over 4 \t}-m^2\t}=\nonumber\\
	&&={1\over 2\pi}\left({m\over2\pi |x-x^\prime| }\right)^{n/2 -1}\,K_{ n/2 -1}\left(m|x-x^\prime|\right)
	\eea
	Particularizing to four dimensions
	\bea
	&&\Delta(x-x^\prime)={1\over 2\pi}{m\over2\pi |x-x^\prime| }\,K_{ 1}\left(m|x-x^\prime|\right)=
{1\over 2\pi}{m\over2\pi |x-x^\prime| }\,\bigg\{{1\over m|x-\xp|}+\nonumber\\
&&+\sum_{k=0}^\infty{(m|x-\xp|)^{1+2k}\over k!(k+1)!}\left(\log\,{m|x-\xp|\over 2}-{\psi(k+1)\over 2}-{\psi(k+2)\over 2}\right)\bigg\}
	\eea
where EulerÁs $\psi$-function is defined by $\psi(x)\equiv {\Gamma (x)^\prime(x)\over \Gamma(x)}$. It can be proved that the only pole of the modified Bessel function $K_1(x)$ on the real axis is located at $x=0$. The singularity structure is exactly the same and is restricted to the coincidence limit. Let us now consider how we can define the integration around it.
	\par	
Consider the one-dimensional integral
\bea
&&I\equiv \int_{-\infty}^\infty {dz}{1\over x-z}f(z){1\over z-\xp}=\text{analytic}+ \int_{x-\e}^{x+\e}dz{f(x)\over x-z}+\int_{\xp-\e}^{\xp+\e}dz{f(\xp)\over z-\xp}=\nonumber\\
&&=\text{analytic}+\pi \left(f(\xp)-f(x)\right)\,i
\eea
The singularity when $x\rightarrow \xp$ will be
\bea
&&I\equiv \int_{-\infty}^\infty {dz}{1\over x-z}f(z){1\over z-x}=\text{analytic}+ \int_{x-\e}^{x+\e}dz{f(z)\over (x-z)^2}=\nonumber\\
&&=\text{analytic}+ \int_{x-\e}^{x+\e}dz{f(x)+f^\prime(x)(x-x)\over (x-z)^2}=\text{analytic}-2 {f(x)\over \e}+\pi i f^\prime(x)
\eea
where
\be
\e=x-\xp
\ee
This means that no extra fancier singularities are created by the convolution process.

\section{Acknowledgements}
 We are grateful to Eduardo Velasco-Aja which collaborated in early stages of this project. We acknowledge partial financial support by the Spanish MINECO through the Centro de excelencia Severo Ochoa Program  under Grant CEX2020-001007-S  funded by MCIN/AEI/10.13039/501100011033.
 
We also acknowledge partial financial support by the Spanish Research Agency (Agencia Estatal de Investigaci\'on) through the grant PID2022-137127NB-I00 funded by MCIN/AEI/10.13039/501100011033/ FEDER, UE

	All authors acknowledge the European Union's Horizon 2020 research and innovation programme under the Marie Sklodowska-Curie grant agreement No 860881-HIDDeN and also byGrant PID2019-108892RB-I00 funded by MCIN/AEI/ 10.13039/501100011033 and by ``ERDF A way of making Europe''.

\appendix

\section{Perturbative expansions of the JNW metric.}
It seems that there has been some confusion in the literature on this topic. We shall try our best to clarify the difgferent limits of the JNW metric

\bea ds^2&&=\left(1-\frac{b}{\r}\right)^{\g}\text{dt}^2-\left(1-\frac{b}{\r}\right)^{-\g}\text{d$\r$}^2-\left(1-\frac{b}{\r}\right)^{1-\g}\r^2\text{d$\Omega $}^2\nonumber\\
\eea
and the scalar field
\be \phi(\r)=\sqrt{1-\g^2\over 2 \kappa^2}\log \left(1-\frac{b}{\r}\right)\ee
remember $b=\frac{r_s}{\g}$, we define
\bea
&&\e_1=\frac{b \g}{\r}\nonumber\\
&&\e_2={1-\g\over \g}
\eea
in this case  we have
\bea ds^2&&=\left[1-\e_1(1+\e_2)\right]^{\frac{1}{1+\e_2}}\text{dt}^2-\left[1-\e_1(1+\e_2)\right]^{-\frac{1}{1+\e_2}}\text{d$\r$}^2\left[1-\e_1(1+\e_2)\right]^{\frac{\e_2}{1+\e_2}}\r^2\text{d$\Omega $}^2\nonumber\\
\eea
and the scalar field
\be \phi(\r)=\sqrt{\frac{1-\frac{1}{(1+\e_2)^2}}{2\kappa^2}}\log \left[1-\e_1(1+\e_2)\right]^{\frac{1}{1+\e_2}}\ee

Consider now in more detail the limits
\bi

\item
First
\bea
&&\e_1\neq 0\nonumber\\
&&\e_2\rightarrow 0\Leftrightarrow \g\rightarrow 1
\eea
at first order, the metric
\bea ds^2&&=\Big[(1-\e_1)+\left(-\e_1+(\e_1-1)\log(1-\e_1)\right)\e_2\Big]\text{dt}^2-\nonumber\\
&&-\Big[\frac{1}{1-\e_1}+\frac{\e_1-(\e_1-1)\log(1-\e_1)}{(\e_1-1)^2}\e_2\Big]\text{d$\r$}^2-\nonumber\\
&&-\Big[1+\e_2\log(1-\e_1)\Big]\r^2\text{d$\Omega$}^2+\mathcal{O}(\e_2^2)
\eea
and the scalar field
\be
\phi(\r)=\sqrt{\frac{1}{\kappa^2}}\log(1-\e_1)\sqrt{e_2}+\mathcal{O}(\e_2^{3/2})
\ee
at order zero, we have the  Schwarzschild metric.
\bea ds^2&&=\left(1-\frac{r_s}{\r}\right)\text{dt}^2-\frac{1}{\left(1-\frac{r_s}{\r}\right)}\text{d$\r$}^2-\r^2\text{d$\Omega$}^2+\mathcal{O}(\e_2^0)
\eea
and the scalar field
\be
\phi(\r)=0+\mathcal{O}(\e_2^0)
\ee
\item
Second
\bea
&&\e_1\rightarrow 0\Leftrightarrow r_s\rightarrow 0\nonumber\\
&&\e_2\neq 0
\eea
at first order, the metric
\bea ds^2&&=\left(1-\e_1\right)\text{dt}^2-\left(1+\e_1\right)\text{d$\r$}^2-\left(1-\e_1\e_2\right)\r^2\text{d$\Omega$}^2+\mathcal{O}(\e_1^1)
\eea
and the scalar field
\be
\phi(\r)=-\sqrt{\frac{\e_2(2+\e_2)}{2\kappa^2}}\e_1+\mathcal{O}(\e_1^1)
\ee
at order zero, we have the  Minkowski metric.
\be ds^2=\text{dt}^2-\text{d$\r$}^2-\r^2\text{d$\Omega$}^2+\mathcal{O}(\e_1^0)
\ee
and the scalar field
\be
\phi(\r)=0+\mathcal{O}(\e_1^0)
\ee

\ei

Finally, if we study the curvature invariants, we obtain.

When $r_s=0$ we have the Minkowski limit and the curvature invariants are
\be
R^2=R_{\m\n}^2=R_{\m\n\r\s}^2=0
\ee
and if $\g=1$ we have the Schwarzschild limit and the curvature invariants are
\bea
&&R^2=R_{\m\n}^2=0\nonumber\\
&&R_{\m\n\r\s}^2=\frac{12r_s^2}{r^6}
\eea

\section{The Plebanski-Demianski solution.}

Plebanski and Demianski \cite{Plebanski} found the general Type D vacuum solution which reads, as written in \cite{Alawadhi}
\be
ds^2={1\over (1-pq)^2}\bigg(2i\left(du+q^2dv\right)dp-2\left(du-p^2 dv\right)dq+{P(p)\over p^2+q^2}\left(du+q^2 dv\right)^2-{Q(q)\over p^2+q^2}\left(du-p^2 dv\right)^2\bigg)
\ee
with
\bea
&P(p)\equiv \g\left(1-p^4\right)+2np-\e p^2+2 m p^3\nonumber\\
&Q(q)\equiv \g\left(1-q^4\right)-2mq+\e q^2-2 n p^3
\eea
and the parameters $(\g,\e,m,n)$ are related to the mass, NUT charge angular momentum and acceleration \cite{Alawadhi}. 
Let us also define the self dual part of the Maxwell two form reads
\be
F^+={m-in\over 2 (p+iq)^2}\bigg(i(du+q^2 dv)dp+(-du+p^2 dv)dq\bigg)
\ee
The {Weyl doubling} \cite{Alawadhi} is a variant of the general doubling construction that represents gravity in terms of Yang-Mills. Then it so happens that Weyl's tensor is  related to the construct $C_{\m\n\r\s}[F^+]$
\be
W_{\m\n\r\s}= \Omega^2 C_{\m\n\r\s}[F^+]
\ee
where
\be
\Omega^2={4i(p+iq)\over (m+in)(1-pq)^4}
\ee
and
\be
C_{\m\n\r\s}[F^+]\equiv {\cal A}\bigg\{F^+_{\m\n}F^+_{\r\s}-F^+_{\r\m}F^+_{\n\s}-{6\over n-2}g_{\m\r} F^+_{\n\l} F_+^\l\,_\s+{3\over (n-1)(n-2)} g_{\m\r }g_{\n \s} F_+^2\bigg\}
\ee
where ${\cal A}$ antisymmetrizes in $[\m\n]$ and $[\r\s]$.

 
\end{document}